\documentclass{vgtc}                          % final (conference style)
\graphicspath{{figures/}{pictures/}{images/}{./}} % where to search for the images
\usepackage{tabularx}
\usepackage{times}
\usepackage{stfloats}% we use Times as the main font
\usepackage{booktabs}
\usepackage{float}
\usepackage{needspace}

\usepackage{mathptmx}                  % use matching math font
\usepackage{amsmath}
\usepackage{amssymb}

\onlineid{0}
\vgtccategory{Short Paper}

\vgtcinsertpkg

\title{Evaluating Human--LLM Alignment on Visualization Preference}

\author{Tanmay Bakshi\thanks{e-mail: t-tbakshi@microsoft.com}\\ %
        \scriptsize Microsoft, India %
\and Chenglong Wang\thanks{e-mail: chenwang@microsoft.com}\\ %
     \scriptsize Microsoft Research, USA
\and Simra Shahid\thanks{e-mail: simrashahid@microsoft.com}\\
     \scriptsize Microsoft, India
}

\abstract{
There are often multiple valid visualizations, spanning different chart types, visual encodings, and aesthetic choices, that can all support a user’s goal of analyzing a dataset. Determining which of these candidate visualizations is most appropriate, however, remains challenging. Prior work has leveraged expert-curated taxonomies to guide visualization design and subsequent quality evaluation, a practical evaluation model must also account for human-centric factors.
Latest visualization tools integrate Large Language Models (LLMs)\footnote{We use the terms Large Language Models (LLMs) and Vision Large Language Models (VLLMs or VLMs) interchangeably.} to generate and refine visualizations, using LLM-as-a-judge to provide design feedback. Without a deeper understanding of whether, and in what ways, human and LLM perceptions of visualizations align, there is a real risk of deploying low-quality agents that do not reflect human judgment.

We study this question through a pilot experiment - we formulate the problem as a preference task over curated chart pairs and compare the judgments of state-of-the-art vision models with those of a panel of human annotators. 
We establish a cautionary tale: such judges risk filtering out informative, compelling displays. We find that frontier models collapse human subjectivity, a rich signal that reveals contrasting design choices, into near-deterministic verdicts.
} % end of abstract

\begin{document}


\firstsection{Introduction}

\maketitle

%% \section{Introduction} %for journal use above \firstsection{..} instead
Visualizations drive the trajectory of business decisions, scientific inquiry, public discourse through journalism, among other fields. For a given dataset and a high-level communicative intent (``a story to tell''), a visualization expert may design multiple valid visualizations--spanning across different chart types, visual encoding and design choices to support them. Determining which candidate visualization is the \emph{most appropriate} (``tells the most compelling story'') constitutes the \textbf{visualization preference task}: a comparative judgment in which a viewer, presented with valid candidate visualizations, chooses a display that serves their need.
A human viewer's perception of a visualization is inherently subjective, not solely measured by presence or absence of information, but rather by how clearly, reliably and efficiently that information is presented, and how it connects to their prior beliefs and task context. Two displays that encode identical data can produce different levels of understanding or decision confidence when presented to the same viewer. Similarly, a single visualization display might be perceived differently across a panel of viewers. Hence, we deal with preference as a \emph{distribution} over objective choices, and not as a binary ground truth.

Visualization literature has largely addressed visualization quality assessment through expert-curated taxonomies of isolated design issues--\emph{truncated axes, misleading ratios, chart-type mismatch etc.} \cite{Cabouat_2025, DBLP:journals/tvcg/LanL25}. While valuable as primers to visualization professionals, these taxonomies act as baseline validity filters. In practice, users frequently choose among charts that have \emph{already cleared this baseline} or are technically ``good enough''. At this juncture, the viewers' latent perception factors drive them to prefer one chart over the other. These factors, rooted in data semantics, aesthetics, etc. cannot be accounted for by static taxonomies. Prior benchmarks that evaluate isolated visualizations against predetermined criteria~\cite{5290695} further miss the comparative nature of human perception: pairwise comparison surfaces preferences that a viewer could not articulate when examining a single chart in isolation~\cite{tufte2001visual}.

Meanwhile, in adjacent AI disciplines, \emph{preference-based signals} have proven to be critical for building reliable agent-based verifiers. From mathematical reasoning, where step-level human preferences guide problem-solving~\cite{Lightman2023LetsVS}, to molecular design where reinforcement-learning agents are steered by preference-aligned scoring functions~\cite{olivecrona2017molecularnovodesigndeep}, the common theme is that once a model's preferences are \emph{aligned} with domain-expert judgement, it can be utilized in a scalable feedback loop that evaluates and improves outputs. This paradigm has already been deployed in data analytics tools such as spreadsheet copilots and dashboard generators; systems of this kind~\cite{10.1145/3706598.3713296, dibia2023lidatoolautomaticgeneration} adopt multi-step agentic pipelines in which a vision-language model (VLM) acts as the preference critic. The quality of every chart surfaced to an end user therefore pivots on the assumption that the critic's preferences reflect human judgment.

Yet, we found that no prior work validates this assumption. VLMs have been benchmarked on Data Viz tasks like chart question-answering~\cite{masry-etal-2022-chartqa} and summarization. However, these benchmarks are centered around the question \emph{``What is presented?''} and not \emph{``How well is it presented?''}. They assess whether a model can \textit{read} a chart, but not whether it can \textit{judge} one---let alone whether its judgments align with those of human viewers.

Through this study, we seek to bridge this gap. We pose the inquiry that : \textbf{When two technically valid charts are constructed to present the same data, do frontier VLMs prefer the same chart that humans prefer?}. We construct an intent-driven, pairwise preference benchmark of 24 diverse chart pairs, and evaluate five vision-language models under stochastic multi-run sampling. By treating annotator disagreements as meaningful signals, we find that there exists a disarray between human and model preferences. 

% , and that the model's ability to recognize \emph{which comparisons are ambiguous for humans} is a sharp differentiator of alignment than overall preference accuracy.

\section{Related Works}
The foundations of visualization quality assessment span grammatical frameworks, decoding-based tasks (judging position, color etc.) ~\cite{Cleveland01091984}, and quantifiable design heuristics such as the data-ink ratio~\cite{tufte2001visual}. Collectively, these works establish principled criteria for visualization design but remain anchored to expert judgment and expensive manual application.

Another relevant line of work probes VLM capabilities in \emph{chart understanding}. Benchmarks such as ChartBench~\cite{xu2024chartbenchbenchmarkcomplexvisual} and
ChartMuseum~\cite{tang2026chartmuseumtestingvisualreasoning} frame visualization comprehension as visual question-answering (VQA), requiring models to extract values, interpret encodings, and infer relationships. Notably, the latter demonstrates how
current models over-rely on textual reasoning and struggle with visually grounded interpretation of implicit encodings and non-trivial layouts---a finding that motivates our investigation of whether these perceptual limitations in VLMs extend to preference judgments and how they relate to human preferences.

On the automation front, a constraint-based recommendation system, \emph{DracoGPT}~\cite{wang2024dracogptextractingvisualizationdesign}, encodes empirical visualization design knowledge as soft and hard rules and checks alignment across 3 LLMs' encoded design practices. VisEval~\cite{chen2024visevalbenchmarkdatavisualization} benchmarks LLM-generated visualizations on pre-determined validity, legality, and readability criteria, while C$^2$~\cite{2koh2025c2scalableautofeedbackllmbased} introduces a novel, LLM-based automated feedback generator that combines static feedback (over chart-type, color, etc.) with dynamic feedback (faithfulness to user intent, etc.) to iteratively improve chart generation. Agentic visualization pipelines such as LIDA \cite{dibia2023lidatoolautomaticgeneration} deploy a code-based LLM evaluator
inside a (generate-evaluate-refine) loop to score and refine charts without human intervention. However, the evaluation signals in all of these systems are either grounded in static, expert-curated taxonomies (axis truncation, label overlap, chart-junk) or remain unvalidated against fine-grained, latent perceptual factors of their critics. This forms the cornerstone of our contribution.

\section{Dataset}
Existing chart evaluation frameworks build primitive visualizations from templates~\cite{Kafle2018DVQAUD} and low-dimensional tables or restrict coverage to narrow domains~\cite{NEURIPS2024_cdf6f8e9} and chart types~\cite{masry-etal-2022-chartqa}. Our data generation pipeline addresses these limitations by modeling visualization construction as a realistic communicative scenario -- integrating diverse, complex datasets, accounting for varying target audience in each domain and clear analytical intent -- resulting in rich, targeted visualizations that exhibit genuine design trade-offs.

\subsection{Data Collection}
To model data visualization scenarios \emph{in the wild}, we used the compiled set of real-world datasets from the ChartUIE-8K Dataset from~\cite{2koh2025c2scalableautofeedbackllmbased}. From 509
source tables, we manually curate 34 datasets spanning 20 thematic categories (education, technology, market research etc.), ensuring broad domain coverage and sufficient structural complexity (3+ data fields each).

\subsection{Data Generation Pipeline}
\noindent\textbf{Stage 1: Structural Annotation} - Each source table is parsed into a structured JSON schema. An LLM annotates every field with its semantic role inferred from co-occurring field names and value types/distributions.

\noindent\textbf{Stage 2: Scenario Construction} - A Chain-of-Thought (CoT) prompt generates 144 scenario briefs across 34 datasets, each specifying a target audience group, data curator's intent and shared context between them. Multiple briefs per dataset explore diverse semantics, and hence diverse communicative framings.

\noindent\textbf{Stage 3: Analytical Question Derivation} - Schemas paired with scenario briefs are fed to an LLM to derive analytical questions that are tractable for visualization, yet trigger interpretive reasoning (compare/correlate/causal analysis etc.). This yields 888 candidate queries, from which 116 questions are curated\footnote{All curation and sampling steps are performed by the first author.} that satisfy two criteria: 
\noindent\textbf{(i)~Analytically non-trivial.}
A question like \textit{``What is the Attack stat of Pikachu?''} is a factual lookup whose answer is independent of chart design. We retain only questions that demand subjective interpretation, such as \textit{``Is there a significant difference in `Attack' between Pok\'{e}mon of different primary types?''}---where the viewer's judgment depends on how the chart encodes the \emph{Attack} statistic.
\noindent\textbf{(ii)~Design-space rich.}
Each query must necessitate utilization of meaningful visualization design choices (using data aggregations, varying chart complexities via layouts, legends, etc.) that affect user perception.

\noindent\textbf{Stage 4: Chart Variant Synthesis} - For each of the 116 queries, an LLM generates 10-15 visualizations as executable \texttt{matplotlib code}, applying distinct transformations and visual encodings to the same data. Via manual curation, we further yield 396 candidate visualizations.

\subsection{Study Set Sampling}
\label{sec:sampling}
From the candidate pool, we jointly sample the tuples (query, candidate A, candidate B) to form the study set. Sampling proceeds in 2 passes-

\noindent\textbf{Coverage Pass: } Queries are thematically coded into intent types\footnote{Derived via open coding of query pool.} (Table~\ref{tab:study-set}) to form a coverage scaffold. Queries are manually sampled to spread the study set across intent types, source domains and chart types. The candidate pool is dominated by comparative questions and certain chart types (box plot, bar chart etc.). This ensures we preserve the rarer intents and the most interesting design trade-offs.

\noindent\textbf{Admissibility Pass: } The sampled charts are required to satisfy a rubric - (i) \textit{Baseline Validity} - Every individual chart must be free of standard taxonomy defects (truncated axes, typographical overlaps etc.) 
(ii) \textit{Design Differentiation} - The two charts must be separable along at least one of structural (chart type, encoding), semantic (data filtration, transforms) or design (annotations, spatial layouts) axes.

\begin{table}[!t]
\centering
\small
\setlength{\tabcolsep}{4pt}
\caption{Thematic composition of the study set}
\label{tab:study-set}
\begin{tabularx}{\columnwidth}{@{}lXr@{}}
\toprule
\textbf{Question type} & \textbf{Example} & \textbf{Count} \\
\midrule
Comparative analysis & How do caffeine levels vary among drink types?         & 6 \\
Filtered comparison  & Among top-ranked countries, how do densities compare?  & 3 \\
Ranking / extrema    & Which stocks have the highest turnover?                & 3 \\
Correlation          & Do properties with more bedrooms have higher prices?   & 3 \\
Anomaly detection    & Are there states with far above-average dog volume?    & 3 \\
Part-to-whole        & Do any genres dominate global sales in specific years? & 3 \\
Temporal trend       & How has global population changed from 1955 to 2024?   & 2 \\
Distribution         & How do undergraduate majors vary among astronauts?     & 1 \\
\bottomrule
\end{tabularx}
\end{table}

\section{USER STUDY}
Prior works in user preference have often relied on \emph{thinking-aloud protocols} etc. to elicit participants' thinking process during interaction with informative artefacts. These approaches assume that users have complete and correct introspective access to the factors that shape their choices. Results in cognitive psychology~\cite{wilson1991thinking} suggest that this methodology degrades task performance quality by shifting their attention to verbalizable but often non-diagnostic features. We posit that this effect shall be especially pronounced for data visualizations---where user perception is rapid and pre-attentive. Our study bypasses the dependence on verbal articulation by eliciting high-level preference signals as an objective choice.

\subsection{Evaluation Dimensions}
To circumvent the problem of burdening annotators with exhaustive visualization taxonomies and to create a scalable annotation platform for future crowd-sourced studies accessible to the common audience, we conducted a structured convergent-design exercise: (i) The authors reviewed literature in visualization research, cognitive science, and human-computer interaction (HCI). (ii) Several candidate evaluation dimensions (grammatical, perceptual and heuristic) were proposed independently and stress-tested through iterative pilot rounds of the annotation task performed on disjoint chart pairs, with the authors serving as trial annotators. (iii) Across multiple iterations, we eliminated ambiguity, extraneous cognitive load, and redundancy and refined candidates until stable inter-rater agreement was reached.
This process yielded three perceptual dimensions, each operationalized as a first-person statement rated on a 5-point Likert scale (Strongly Disagree to Strongly Agree):
\begin{itemize}
    \item \textbf{Readability} --- \emph{``I think the chart is easy to
    understand at a glance.''}
    \item \textbf{Precision} --- \emph{``I think the chart answers the given
    question.''}
    \item \textbf{Aesthetics} --- \emph{``I think the chart is visually
    appealing and attractive.''}
\end{itemize}
\noindent We do not claim these dimensions to be theoretically orthogonal; rather, they represent the minimal set on which the authors could reach stable, independent agreement. 
% These dimensions' results have not been reported further. Establish that these are just to ground the annotator's understanding of the task, and aid him in making decision
Crucially, within the scope of this work the Likert ratings are not themselves an object of analysis, but serve an instrumental role. They act as lightweight scaffolding that anchors annotators to the salient facets of the task---readability, precision, and aesthetics. While these scores are rich signals in themselves, we leave the analysis of these dimensions to future work.

\subsection{Procedure}
% We recruit a panel of 5 volunteers (to avoid injecting bias, none of the authors are included as participants) to participate in the study. We curate a subset of 22 unique analytical questions drawn from 17 charts spanning 4 domains: Business/Market, Public Policy/Governance, Science/Education, and Healthcare/Insurance. For 2 of these questions, two visualization pairs are curated, resulting in a total of 24 data points.
We recruit a panel of 10 volunteers; to avoid injecting authorial bias, none of the authors participate in the study. Each participant is briefed on the task objectives and evaluation dimensions in a synchronous session, after which they complete the task independently at their own pace. Annotations are collected through a custom-built interface, iteratively refined based on pilot feedback to minimize ambiguity and cognitive load during the annotation workflow.\footnote{A walkthrough of the annotation interface is included in the supplementary material.}
Each participant first completes an objective, pre-study questionnaire assessing their experience with data handling, their familiarity with visualizations and their self-reported confidence in interpreting data visualizations.

Each of the 24 pairs is then presented in a two-phase protocol: 
\begin{enumerate}
    \item \textbf{Independent evaluation: } Chart~A and Chart~B are shown \emph{sequentially} alongside the analytical question. For each chart in isolation, the participant rates Readability, Precision, and Aesthetics on Likert scales. This phase establishes per-display context rooted in the question and invites
    analytical engagement before making comparative judgments.
    \item \textbf{Pairwise preference: } After sequential rating, both charts are displayed \emph{side-by-side} together with the original analytical question.
    The participant selects one of three options: \emph{Chart~A}, \emph{Chart~B}, or \emph{Equal Preference}---a three-bin choice that captures objective display preference from the user.
\end{enumerate}
\noindent \textbf{Note: } The presentation order of charts was randomized for the participants and the models.
\noindent Each phase allows participants to optionally add comments about their feelings during the task. \footnote{these comments are not a part of our findings in this work.}Following the annotation exercise, participants complete a post-study questionnaire that gauges task engagement and fatigue, which is subsequently used as a control over response quality. The observed \textbf{mean engagement time} was 148.1 seconds per pair.

\subsection{Participant Description}
The study included 10 participants, out of which, 7 hold bachelor's degrees, 2 hold a masters degree and 1 holds a PhD degree. Most participants reported a professional experience between 1-5 years (including internships etc.), while 1 participant reports less than 1 year of professional experience.
The participants were all frequent consumers of data visualizations, reporting high degree of confidence in their ability to perceive data visualizations (90\%), interpretation of visualizations with multiple encodings (70\%), and handling unconventional visualization types (70\%).
\begin{figure}[tbp]
    \centering
    \includegraphics[width=\columnwidth]{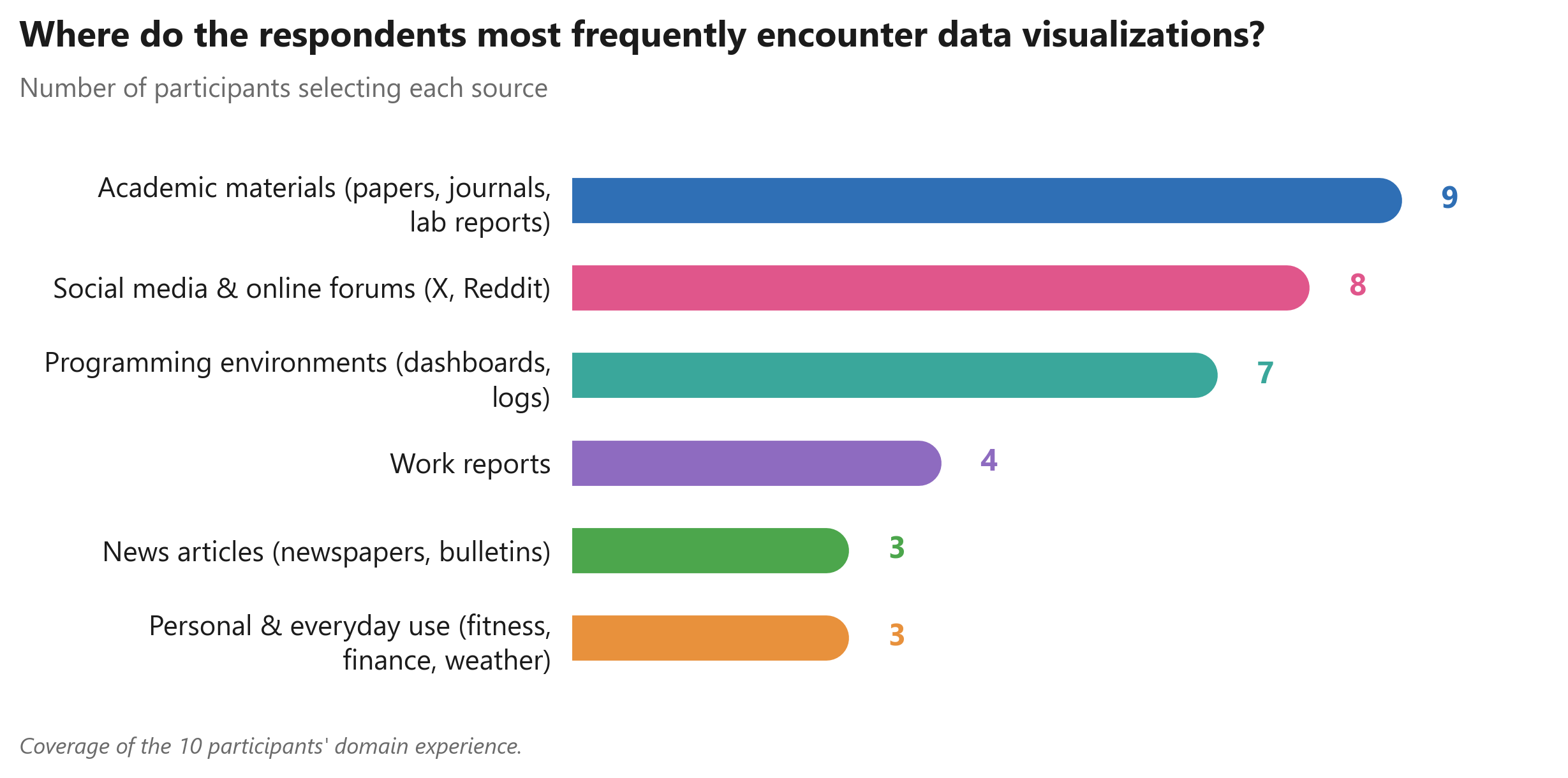}
    \caption{Reported familiarity domains by participants}
    \label{fig:source-counts}
\end{figure}
Our panel is visualization-literate, identifying with routine users of specialized spreadsheet AI tools, and hence form an appropriate population set for this study.

\section{Evaluation}
% \label{sec:evaluation}
Subjective preferences have no unique gold label. We treat participant disagreement as signal~\cite{Aroyo2015TruthIA}. 
For each chart pair $i$, ten human votes define the empirical distribution
$\mathbf{h}_i=(h_{i,A},h_{i,B},h_{i,\mathrm{Equal}})$
over $\{A,B,\mathrm{Equal}\}$; analogously, model judgments with stochastic sampling
($T{=}0.75$) define an empirical distribution $\mathbf{m}_i$~\cite{kuhn2023semanticuncertaintylinguisticinvariances}.
We evaluate Claude Opus~4.5, Claude Sonnet~4.5, GPT-5~Chat, GPT-4V, and
Phi-4 Multimodal under identical conditions and prompt scaffold: analytical question, chart images, and evaluation dimensions---eliciting Likert ratings for each chart, a three-way choice and corresponding rationale.\footnote{Prompts are provided in the supplement.} 

\subsection{Metrics}
\label{sec:metrics}
Asking whether a model ``agrees with humans'' bundles two separable
capabilities: whether it \emph{picks} what the panel picks (agreement), and whether it
\emph{knows how much the panel disagrees} (calibration). We measure each: Jensen--Shannon divergence (JSD) for agreement, and a paired entropy test for calibration. We report the following statistics: 
\begin{equation}
J=\frac{1}{P}\sum_{i=1}^{P}\mathrm{JSD}(\mathbf{h}_i,\mathbf{m}_i),
\qquad
T=\operatorname{median}_{i}\!\left[H(\mathbf{h}_i)-H(\mathbf{m}_i)\right].
\label{eq:metrics}
\end{equation}
JSD is a symmetric, information-theoretic distance between two distributions. The agreement statistic $J$ is the mean over $P=24$ pairs. $J\in[0,1]$ measures overall distributional disagreement between the human panel and each model (lower is better). Let $H(\cdot)$ be Shannon entropy in bits. The per-pair sharpness excess is $d_i = H(\mathbf{h}_i) - H(\mathbf{m}_i)$, and the calibration statistic $T$ is its median over the study set, in bits. $T>0$ means the model is typically more certain than the panel, and $T<0$ means it is typically less certain. $J$ identifies whether the distributions differ, while $T$ helps interpret whether excess confidence contributes to that difference. We also report $n_+$, the number of pairs for which $d_i > 0$, i.e.\ the number of pairs on which the model exhibited more certainty across runs than the human panel.

\subsection{Calibrated Reference}
\label{sec:calref}
Zero is not the right baseline: a perfectly calibrated predictor sampled only
$K{=}10$ times is itself noisy. To locate that floor we place a symmetric
$\mathrm{Dirichlet}(1)$ prior on each pair's vote split and, for each of
$B{=}2000$ replicates, draw that split from the posterior before sampling $K$
votes from it. Drawing the split rather than fixing it at the posterior mean
lets the reference inherit the panel's own estimation uncertainty. Built from
the observed votes alone, this reference---not zero---is what the model rows of
Table~\ref{tab:results} should be read against.\footnote{The supplement reports
the corresponding tail probabilities and their multiplicity adjustment.}

\section{Observations}
\label{sec:results}
Differences \emph{between} individual models are not resolvable at $P{=}24$ and we make no claim about model rankings; we compare frontier models as a class against the calibrated reference. The $J$ values show that at the data population level, the model evaluators' agreement sits roughly 4 times as far from the human panel as a calibrated predictor's. This confidence effect is universal, as frontier VLMs appear more certain than the panel on $21$--$24$ of the $24$ pairs, whereas a calibrated predictor is more certain on $11$ pairs on average and on at most $15$, representing an evident overconfidence. The models don't merely disagree with the panel, as the $J$ values show, they also collapse genuine human disagreement into deterministic verdicts. 
Phi-4 is the only model at the calibration floor. However, on closer inspection, it is revealed that in $57\%$ of its total responses, it systematically misidentifies chart types (box plots as scatter plots, violin plots as bar charts etc.) or fails to name one, violating the ``visualization literacy'' confound. Hence, we attribute its performance to instability.
Our results are robust as this trend survives order randomization and prompt framing effects\footnote{Supplementary material}.
% Its rationales appeal to
% orientation, spacing and colour contrast rather than to encoding or task fit

\begin{table}[htbp]
\centering
\setlength{\tabcolsep}{2pt}
% booktabs rule padding is what breaks vertical rules; zero it locally.
\setlength{\aboverulesep}{0pt}
\setlength{\belowrulesep}{0pt}
\newcommand{\ci}[2]{{\scriptsize\,$[#1,\,#2]$}}
\caption{Agreement and calibration relative to the calibrated reference}
\label{tab:results}
\begin{tabular}{l|c|cc}
\toprule
& \multicolumn{1}{c|}{\textbf{Agreement}} & \multicolumn{2}{c}{\textbf{Calibration}} \\
\textbf{Predictor} & $J\downarrow$ & $T\downarrow$ & $n_+\downarrow$ \\
\midrule
\emph{Calibrated reference} & $0.08$\ci{0.05}{0.11} & $0.02$\ci{-0.11}{0.20} & $11$\ci{7}{15} \\
\midrule
GPT-4V            & $0.29$\ci{0.21}{0.38} & $0.90$ & $21$ \\
GPT-5 Chat        & $0.29$\ci{0.21}{0.38} & $0.90$ & $21$ \\
Claude Opus 4.5   & $0.29$\ci{0.21}{0.38} & $0.88$ & $24$ \\
Claude Sonnet 4.5 & $0.31$\ci{0.22}{0.39} & $0.91$ & $24$ \\
Phi-4 Multimodal  & $0.16$\ci{0.12}{0.21} & $0.19$ & $15$ \\
\bottomrule
\end{tabular}
\end{table}

\section{Case Studies}
\label{sec:cases}
We present 2 example cases from the study -- one where humans and models converge, the other where they systematically diverge:
\newcommand{\caseheading}[1]{%
    \par\addvspace{0.35\baselineskip}%
    \noindent\textbf{#1}\par\nobreak\vspace{2pt}%
}

\begin{figure}[htbp]
    \centering
    \includegraphics[width=0.9\columnwidth]{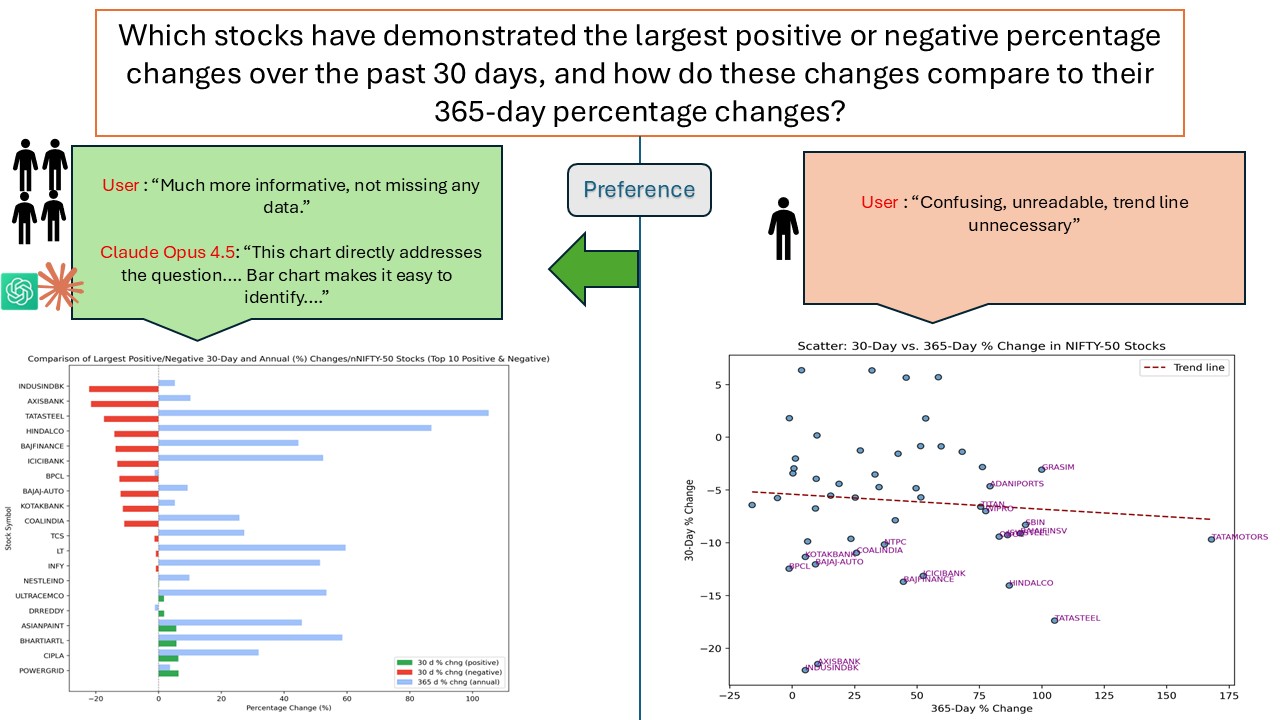}
    \caption{Clear agreement between models and participants: (i) Chart A
    (left), a diverging bar chart; and (ii) Chart B (right), a scatter plot.}
    \label{fig:agreement}
\end{figure}

\caseheading{Case I: Agreement on Stock Performance Charts}
Only 1 participant chose Chart~B\@ as shown in Figure~\ref{fig:agreement}, citing its completeness for their choice
(``much more informative, not missing any data''),
while others cited clutter (``Unreadable,'' ``Cluttered'') and the irrelevance of Chart~B's trend line for not picking it. The four frontier models unanimously selected Chart~A in every run (40/40), strongly aligning with human preferences. The models exhibit a consistent rationale centered around \emph{question--chart semantic congruence}. The question asks ``which stocks'' (requires explicit identification) and ``how do these compare'' (requires comparative encoding). Chart~A's ranked bar layout directly supports both:
stocks are named, sorted by 30-day magnitude, and their 365-day changes are shown side-by-side. Opus 4.5 describes it as ``directly addressing the question by showing the top 10 positive and negative 30-day performers.''

\begin{figure}[htbp]
    \centering
    \includegraphics[width=0.9\columnwidth]{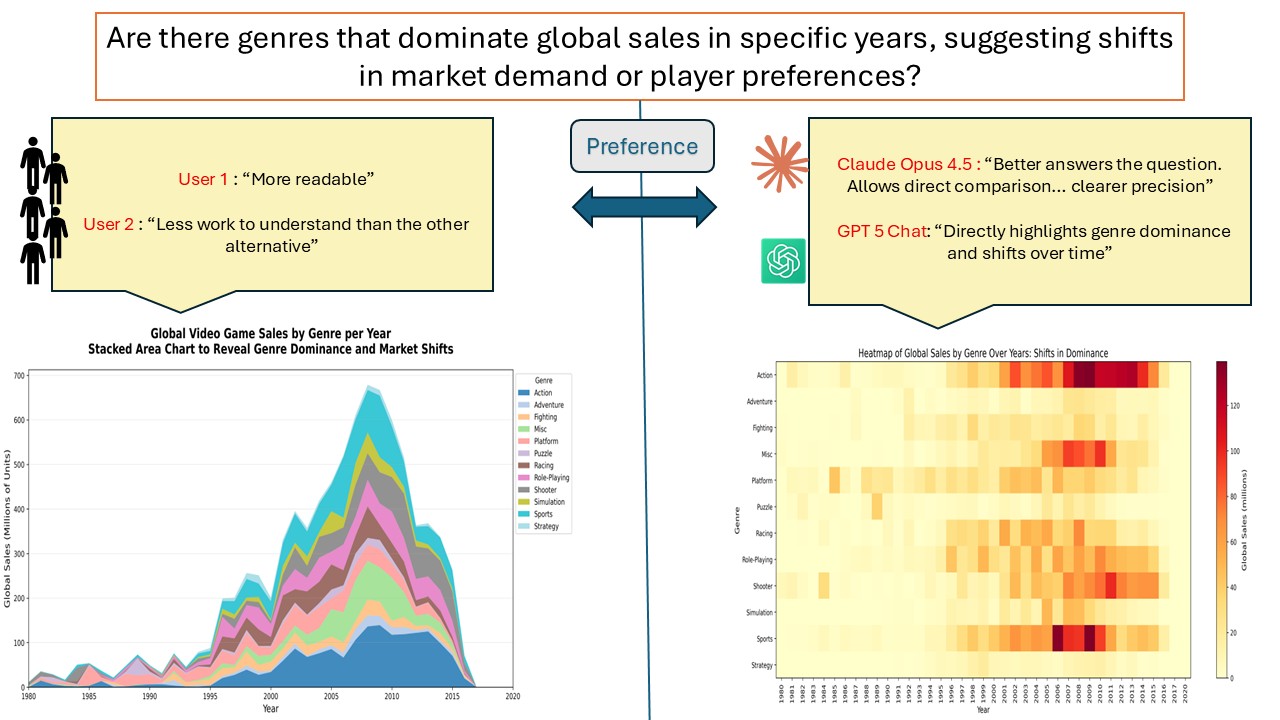}
    \caption{Disagreement: (i) Chart A (left),
    a stacked area chart; and (ii) Chart B (right), a heatmap.}
    \label{fig:disagreement}
\end{figure}

\caseheading{Case II: Disagreement on Video Game Sales Charts}
In Figure~\ref{fig:disagreement}, 9 of the 10 participants chose Chart~A, while the models overwhelmingly selected Chart~B: Claude Opus~4.5 and Claude Sonnet~4.5 in all 10 runs, and GPT-4V and GPT-5~Chat in 6 and 8 runs respectively.
Claude Opus 4.5 explains how ``the heatmap allows immediate identification of which genres dominated and how preferences shifted over time''. Whereas the human annotators are overburdened by the high cognitive load associated with parsing the heatmap, and instead, prefer the intuitive outlook provided by the stacked area chart, despite the fact that it does not answer the question as precisely (via year-by-year breakdown) as the heatmap. The models seem to perform a highly nuanced, analytical breakdown of the task and select the chart that maximizes precision. Instead, humans favor the communicative aspect that is perceived as ``unambiguous'' and tend to avoid complex analysis, revealing fundamental \emph{Cognitive-Load asymmetry}.

\section{Conclusion}
These cases mechanise the aggregate numbers. In Case~I, agreement arises when the task's semantic structure (ranking, comparison) maps unambiguously to one chart type, and both humans and models recognize this. In Case~II, disagreement emerges when the question allows for two valid interpretive strategies: \emph{narrative appeal} vs \emph{analytical decomposition}. This aspect is inherently invisible in model judgements. Models consistently optimize for ``sheer answerability'' of the question, hinging  upon factors like ``chart-type appropriateness'' to match the query etc., while humans optimize for the visualization that minimizes cognitive load and tells them a convincing story. 
We do not position these metrics to evaluate visualization judges, but as diagnostic tools: sufficient to establish that a genuine gap exists in accounting for rich, human-centric preferences that a visualization agent would not surface. To truly leverage the power of agentic visualization tools, a \emph{human-centric} framework that integrates subjective design choices and elements of perceptual fluency rather than pure analytical rigor is needed, and subjective human preference should be advocated as a first-class citizen in chart quality evaluation. We call for a broader inquiry via crowdsourced studies and preference-aware post-training objectives.

\section*{Supplemental Materials}
\label{sec:supplemental_materials}
Supplemental materials are provided under an MIT
license. They contain the data pipeline prompts, preference prompts, UI, model and anonymized annotator responses, and the quantitative results to support the claims in this paper.

% \section*{Figure Credits}
% \label{sec:figure_credits}

% Refer to the instructions for this section (\cref{sec:figure_credits_inst}).
% Here are the actual figure credits for this template:

% \Cref{fig:teaser} image credit: Scott Miller / Special to the Vancouver Sun, January 22, 2009, page A6.

% \Cref{fig:vis_papers} is a partial recreation of Fig.\ 1 from \cite{Isenberg:2017:VMC}, which is in the public domain.

% %% if specified like this the section will be committed in review mode
% The authors wish to thank A, B, and C. This work was supported in part by
% a grant from XYZ.}

%\bibliographystyle{abbrv}
%\bibliographystyle{abbrv-doi}
\bibliographystyle{abbrv-doi-narrow}

\bibliography{template}

@article{Cleveland01091984,
author = {William S. Cleveland and Robert McGill},
title = {Graphical Perception: Theory, Experimentation, and Application to the Development of Graphical Methods},
journal = {Journal of the American Statistical Association},
volume = {79},
number = {387},
pages = {531--554},
year = {1984},
publisher = {Taylor \& Francis},
doi = {10.1080/01621459.1984.10478080},
URL = {
        https://doi.org/10.1080/01621459.1984.10478080
},
eprint = { 
    
        https://doi.org/10.1080/01621459.1984.10478080
}
}

@article{Cabouat_2025,
   title={PREVis: Perceived Readability Evaluation for Visualizations},
   volume={31},
   ISSN={2160-9306},
   url={http://dx.doi.org/10.1109/TVCG.2024.3456318},
   DOI={10.1109/tvcg.2024.3456318},
   number={1},
   journal={IEEE Transactions on Visualization and Computer Graphics},
   publisher={Institute of Electrical and Electronics Engineers (IEEE)},
   author={Cabouat, Anne-Flore and He, Tingying and Isenberg, Petra and Isenberg, Tobias},
   year={2025},
   month=Jan, pages={1083–1093} }

@misc{2koh2025c2scalableautofeedbackllmbased,
      title={$C^2$: Scalable Auto-Feedback for LLM-based Chart Generation}, 
      author={Woosung Koh and Jang Han Yoon and MinHyung Lee and Youngjin Song and Jaegwan Cho and Jaehyun Kang and Taehyeon Kim and Se-Young Yun and Youngjae Yu and Bongshin Lee},
      year={2025},
      eprint={2410.18652},
      archivePrefix={arXiv},
      primaryClass={cs.LG},
      url={https://arxiv.org/abs/2410.18652}, 
}

@misc{wang2024dracogptextractingvisualizationdesign,
      title={DracoGPT: Extracting Visualization Design Preferences from Large Language Models}, 
      author={Huichen Will Wang and Mitchell Gordon and Leilani Battle and Jeffrey Heer},
      year={2024},
      eprint={2408.06845},
      archivePrefix={arXiv},
      primaryClass={cs.HC},
      url={https://arxiv.org/abs/2408.06845}, 
}

@article{DBLP:journals/tvcg/LanL25,
  author       = {Xingyu Lan and
                  Yu Liu},
  title        = {"I Came Across a Junk": Understanding Design Flaws of Data
                  Visualization from the Public's Perspective},
  journal      = {{IEEE} Trans. Vis. Comput. Graph.},
  volume       = {31},
  number       = {1},
  pages        = {393--403},
  year         = {2025},
  url          = {https://doi.org/10.1109/TVCG.2024.3456341},
  doi          = {10.1109/TVCG.2024.3456341},
  bibsource    = {dblp computer science bibliography, https://dblp.org}
}

@ARTICLE{5290695,
  author={Munzner, Tamara},
  journal={IEEE Transactions on Visualization and Computer Graphics}, 
  title={A Nested Model for Visualization Design and Validation}, 
  year={2009},
  volume={15},
  number={6},
  pages={921-928},
  doi={10.1109/TVCG.2009.111}}

@book{tufte2001visual,
  title={The Visual Display of Quantitative Information},
  author={Tufte, E.R.},
  isbn={9780961392147},
  lccn={20271866},
  url={https://books.google.co.in/books?id=GTd5oQEACAAJ},
  year={2001},
  publisher={Graphics Press}
}

@inproceedings{masry-etal-2022-chartqa,
    title = "{C}hart{QA}: A Benchmark for Question Answering about Charts with Visual and Logical Reasoning",
    author = "Masry, Ahmed  and
      Long, Do Xuan  and
      Tan, Jia Qing  and
      Joty, Shafiq  and
      Hoque, Enamul",
    editor = "Muresan, Smaranda  and
      Nakov, Preslav  and
      Villavicencio, Aline",
    booktitle = "Findings of the Association for Computational Linguistics: ACL 2022",
    month = may,
    year = "2022",
    address = "Dublin, Ireland",
    publisher = "Association for Computational Linguistics",
    url = "https://aclanthology.org/2022.findings-acl.177/",
    doi = "10.18653/v1/2022.findings-acl.177",
    pages = "2263--2279"
}

@inproceedings{NEURIPS2024_cdf6f8e9,
 author = {Wang, Zirui and Xia, Mengzhou and He, Luxi and Chen, Howard and Liu, Yitao and Zhu, Richard and Liang, Kaiqu and Wu, Xindi and Liu, Haotian and Malladi, Sadhika and Chevalier, Alexis and Arora, Sanjeev and Chen, Danqi},
 booktitle = {Advances in Neural Information Processing Systems},
 doi = {10.52202/079017-3609},
 editor = {A. Globerson and L. Mackey and D. Belgrave and A. Fan and U. Paquet and J. Tomczak and C. Zhang},
 pages = {113569--113697},
 publisher = {Curran Associates, Inc.},
 title = {CharXiv: Charting Gaps in Realistic Chart Understanding in Multimodal LLMs},
 url = {https://proceedings.neurips.cc/paper_files/paper/2024/file/cdf6f8e9fd9aeaf79b6024caec24f15b-Paper-Datasets_and_Benchmarks_Track.pdf},
 volume = {37},
 year = {2024}
}

@misc{xu2024chartbenchbenchmarkcomplexvisual,
      title={ChartBench: A Benchmark for Complex Visual Reasoning in Charts}, 
      author={Zhengzhuo Xu and Sinan Du and Yiyan Qi and Chengjin Xu and Chun Yuan and Jian Guo},
      year={2024},
      eprint={2312.15915},
      archivePrefix={arXiv},
      primaryClass={cs.CV},
      url={https://arxiv.org/abs/2312.15915}, 
}

@misc{tang2026chartmuseumtestingvisualreasoning,
      title={ChartMuseum: Testing Visual Reasoning Capabilities of Large Vision-Language Models}, 
      author={Liyan Tang and Grace Kim and Xinyu Zhao and Thom Lake and Wenxuan Ding and Fangcong Yin and Prasann Singhal and Manya Wadhwa and Zeyu Leo Liu and Zayne Sprague and Ramya Namuduri and Bodun Hu and Juan Diego Rodriguez and Puyuan Peng and Greg Durrett},
      year={2026},
      eprint={2505.13444},
      archivePrefix={arXiv},
      primaryClass={cs.CL},
      url={https://arxiv.org/abs/2505.13444}, 
}

@misc{chen2024visevalbenchmarkdatavisualization,
      title={VisEval: A Benchmark for Data Visualization in the Era of Large Language Models}, 
      author={Nan Chen and Yuge Zhang and Jiahang Xu and Kan Ren and Yuqing Yang},
      year={2024},
      eprint={2407.00981},
      archivePrefix={arXiv},
      primaryClass={cs.HC},
      url={https://arxiv.org/abs/2407.00981}, 
}

@article{wilson1991thinking,
author = {Wilson, Timothy and Schooler, Jonathan},
year = {1991},
month = {03},
pages = {181-92},
title = {Thinking Too Much: Introspection Can Reduce the Quality of Preferences and Decisions},
volume = {60},
journal = {Journal of personality and social psychology},
doi = {10.1037//0022-3514.60.2.181}
}

@article{Aroyo2015TruthIA,
  title={Truth Is a Lie: Crowd Truth and the Seven Myths of Human Annotation},
  author={Lora Aroyo and Chris Welty},
  journal={AI Mag.},
  year={2015},
  volume={36},
  pages={15-24},
  url={https://api.semanticscholar.org/CorpusID:6134326}
}

@article{Lightman2023LetsVS,
  title={Let's Verify Step by Step},
  author={Hunter Lightman and Vineet Kosaraju and Yura Burda and Harrison Edwards and Bowen Baker and Teddy Lee and Jan Leike and John Schulman and Ilya Sutskever and Karl Cobbe},
  journal={ArXiv},
  year={2023},
  volume={abs/2305.20050},
  url={https://api.semanticscholar.org/CorpusID:258987659}
}

@misc{olivecrona2017molecularnovodesigndeep,
      title={Molecular De Novo Design through Deep Reinforcement Learning}, 
      author={Marcus Olivecrona and Thomas Blaschke and Ola Engkvist and Hongming Chen},
      year={2017},
      eprint={1704.07555},
      archivePrefix={arXiv},
      primaryClass={cs.AI},
      url={https://arxiv.org/abs/1704.07555}, 
}

@misc{dibia2023lidatoolautomaticgeneration,
      title={LIDA: A Tool for Automatic Generation of Grammar-Agnostic Visualizations and Infographics using Large Language Models}, 
      author={Victor Dibia},
      year={2023},
      eprint={2303.02927},
      archivePrefix={arXiv},
      primaryClass={cs.AI},
      url={https://arxiv.org/abs/2303.02927}, 
}

@inproceedings{10.1145/3706598.3713296,
author = {Wang, Chenglong and Lee, Bongshin and Drucker, Steven M. and Marshall, Dan and Gao, Jianfeng},
title = {Data Formulator 2: Iterative Creation of Data Visualizations, with AI Transforming Data Along the Way},
year = {2025},
isbn = {9798400713941},
publisher = {Association for Computing Machinery},
address = {New York, NY, USA},
url = {https://doi.org/10.1145/3706598.3713296},
doi = {10.1145/3706598.3713296},
booktitle = {Proceedings of the 2025 CHI Conference on Human Factors in Computing Systems},
articleno = {677},
numpages = {17},
location = {
},
series = {CHI '25}
}

@misc{kuhn2023semanticuncertaintylinguisticinvariances,
      title={Semantic Uncertainty: Linguistic Invariances for Uncertainty Estimation in Natural Language Generation}, 
      author={Lorenz Kuhn and Yarin Gal and Sebastian Farquhar},
      year={2023},
      eprint={2302.09664},
      archivePrefix={arXiv},
      primaryClass={cs.CL},
      url={https://arxiv.org/abs/2302.09664}, 
}

@article{Kafle2018DVQAUD,
  title={DVQA: Understanding Data Visualizations via Question Answering},
  author={Kushal Kafle and Scott D. Cohen and Brian L. Price and Christopher Kanan},
  journal={2018 IEEE/CVF Conference on Computer Vision and Pattern Recognition},
  year={2018},
  pages={5648-5656},
  url={https://api.semanticscholar.org/CorpusID:4445015}
}
\end{document}